\documentclass[twocolumn,showpacs,prl,10pt]{revtex4}
\usepackage[english]{babel}
\usepackage{amsmath,amssymb} 
\usepackage{times}
\usepackage{epsfig} 

\begin{document} 
%\draft
\title{Temperature and doping dependence of high-energy kink in
cuprates}

\author{M. M. Zemlji\v c$^{1}$, P. Prelov\v sek$^{1,2}$ and
T. Tohyama$^{3}$}
\affiliation{$^1$J.\ Stefan Institute, SI-1000 Ljubljana, Slovenia}
\affiliation{$^2$ Faculty of Mathematics and Physics, University of
Ljubljana, SI-1000 Ljubljana, Slovenia}
\affiliation{$^3$ Yukawa Institute for Theoretical Physics,
Kyoto University, Kyoto 606-8502, Japan}

\date{\today}
                   
\begin{abstract}
It is shown that spectral functions within the extended $t$-$J$ model,
evaluated using the finite-temperature diagonalization of small
clusters, exhibit the high-energy kink in single-particle dispersion
consistent with recent angle-resolved photoemission results on
hole-doped cuprates. The kink and waterfall-like features persist up
to large doping and to temperatures beyond $J$ hence the origin can be
generally attributed to strong correlations and incoherent hole
propagation at large binding energies. In contrast, our analysis
predicts that electron-doped cuprates do not exhibit these phenomena
in photoemission.
 
\end{abstract}

\maketitle 
 
The anomalous properties of quasiparticles (QP) in cuprates are most
directly probed by the angle-resolved photoemission spectroscopy
(ARPES) \cite{dama}. Besides the Fermi surface development, pseudogap
features and low-energy kink \cite{dama} considerable attention has
been recently devoted to the high-energy kink (HEK) observed by ARPES
quite universally in hole-doped cuprates 
\cite{ronn,graf,xie,vall,pan,chan,park,meev,inos}.
The anomaly appears in the QP dispersion along the zone diagonal $(0,0)$ -
$(\pi/2,\pi/2)$ as a kink at binding energies typically $E_1 \sim
0.4$~eV followed by a fast drop - 'waterfall' to the high-energy scale $E_2 \sim
1$~eV being well pronounced around the zone center ${\bf k} \sim (0,0)$. The
HEK seems to exist in a broad range of hole-doped cuprates: from
undoped \cite{ronn,meev}, underdoped \cite{graf,vall,meev}, optimally
doped \cite{graf,vall,chan,meev}, overdoped
\cite{graf,pan,meev,inos} to highly overdoped regime
\cite{xie,meev}. Some ARPES spectra \cite{graf,pan} give indications for
the coexistence of two branches in the part of the Brillouin zone: one
representing the renormalized QP band reaching $E \sim E_1$ and the
second being the remnant of the high-energy unrenormalized band at $E
\sim E_2$. Analogous evidence for the HEK in electron-doped cuprates
\cite{park,pan} is less conclusive.

Theoretically, the origin of the HEK is presently lively debated.
Since in contrast to low-energy kink \cite{dama} the energy $E_1$ is
too high to be attributed to phonons, several aspects of strong
correlations are given as a possible explanation. The similarity to spectral
functions of one-dimensional (1D) chain cuprate SrCuO$_2$ \cite{kim}
with pronounced two component spectra, i.e., spinon and holon
branches, seems to support the long-sought spinon-holon scenario also
for two-dimensional (2D) cuprates \cite{graf}. On the other hand,
alternative explanations with string excitations of a QP in an
antiferromagnet (AFM) \cite{mano}, split QP band within the slave-boson
theory \cite{wang}, and the vicinity to a quantum critical point
\cite{zhu} are not restricted to 1D. Recent numerical calculations within
the Hubbard model support the existence of the HEK in prototype models of
correlated electrons, both for the undoped system \cite{bycz} as well
as in the large-doping regime \cite{macr}, where the origin of HEK is
attributed to high-energy spin correlations \cite{macr,marw}.

In the following we present finite-temperature numerical results
within the prototype $t$-$J$ model of strongly correlated electrons in
cuprates.  They reveal the existence of the HEK in a broad range of
hole concentration $c_h$ and temperature $T$ in the $\omega<0$ sector
of spectral functions $A({\bf k},\omega)$, corresponding to ARPES in
hole-doped cuprates. Well pronounced at intermediate and large doping
as the waterfall-like dispersion, the HEK develops at lower doping and $T<J$
into two partly coexisting branches, the renormalized QP band and a broad
bottom band. An important fact for the interpretation is the observed
persistence of the HEK up to high $T \sim t >J$ which gives strong support to the
scenario that the HEK and waterfall are quite universal signatures of
strong correlations and only indirectly connected to low-$T$ phenomena
as the longer-range AFM and superconductivity in these
materials. Also, strong asymmetry in $\omega$ leads to the conclusion that
analogous phenomena in electron-doped cuprates should be absent within
ARPES spectra.

We study the single-particle excitations within the extended $t$-$J$
model
\begin{equation}
H=-\sum_{i,j,s}t_{ij}\tilde{c}^\dagger_{js}\tilde{c}_{is}
+J\sum_{\langle ij\rangle}{\bf S}_i\cdot {\bf S}_j, \label{eq1} 
\end{equation}
where $\tilde{c}^\dagger_{is}$ are projected fermionic operators not
allowing for the double occupancy of sites.  As relevant for cuprates
we consider the model on a square lattice and include besides the
nearest-neighbor $t_{ij}=t$ also the second-neighbor $t_{ij}=t^\prime$
and the third-neighbor hopping $t_{ij}=t^{\prime\prime}$. We present
in the following results for
$t^\prime=-0.25~t$, $t^{\prime\prime}=0.12~t$, $J=0.4~t$
\cite{tohy1,tohy2} chosen to reproduce well properties of 
hole-doped cuprates, e.g., the measured Fermi surface.

We calculate the spectral function $A({\bf k},\omega)$ using the usual
$T=0$ exact diagonalization method and the
finite-temperature Lanczos method (FTLM) for $T>0$ \cite{jprev}. Systems
considered here are tilted square lattices of $N=18,20$ sites with
finite concentration of holes $c_h=N_h/N$ doped into the reference
undoped AFM insulator. Since fixed boundary conditions on small systems
allow only a discrete set of wavevectors
${\bf k}_l, l=1,N$ we employ twisted boundary conditions to scan the
whole Brillouin zone \cite{tohy2}, ${\bf k}={\bf k}_l+\vec \theta$ by
introducing hopping elements $t_{ij} \to \tilde t_{ij} = t_{ij}
~\mathrm {exp}(i\vec{\theta}\cdot\vec{r}_{ij})$ in Eq.~(\ref{eq1}).
For details of the application of the FTLM to spectral functions we
refer to Ref.\cite{zeml1}. Besides the evident possibility of obtaining $T>0$
results the FTLM allows for a reliable evaluation
of the self energy $\Sigma({\bf k}, \omega)$ which is essential for the
interpretation of observed phenomena.

First we present results for $A({\bf k},\omega)$ at
hole-doping $c_h=0.1$ corresponding to underdoped regime calculated on
a system of $N=20$ sites. In Fig.~1 we
present the weight map of $A({\bf k},\omega)$ along the diagonal and the
edge directions within the first Brillouin zone and its evolution with
increasing $T$. The $T=0$ result in Fig.~1a is obtained by the ground state
Lanczos procedure \cite{tohy2}, while in Fig.~1b,c,d FTLM results are
shown for increasing $T/t=0.2, 0.4, 0.75$.

\begin{figure}[htb]  
\centering 
\epsfig{file=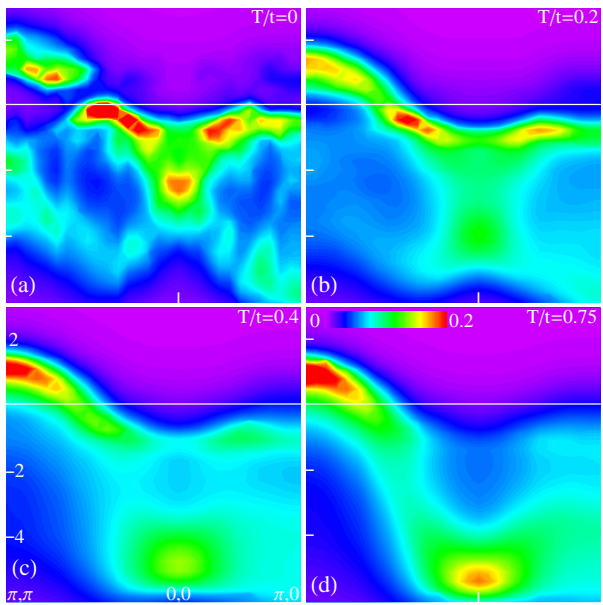,width=85mm,angle=0} 
\caption{(Color online) Weight map of $A({\bf k},\omega)$ vs. ${\bf k}$ 
along symmetry lines in the Brillouin zone for $c_h=0.1$ and different
$T/t$.}
\label{fig1}   
\end{figure} 

If one concentrates on the dispersion along the zone diagonal
$(0,0)-(\pi,\pi)$ it is easy to recognize the HEK feature at $\omega\sim -t$ for all
presented $T$. In fact, a pronounced waterfall-like single band
dispersion is evident even at very high $T\sim t > J$, where the steep drop appears close
to ${\bf k} \sim x(\pi,\pi)$ with $x \sim 0.3$. It should be, however, noted that
$T \sim t$ represents already very high $T$ in this doping regime which leads to
a substantial shift of the chemical potential so that the Fermi surface is
tending towards $(\pi,\pi)$ as evident in Fig.~1d.  

With lowering $T<J$ the $\omega=0$ (Fermi surface) crossing of the
dispersion along the zone diagonal approaches ${\bf k} \sim (\pi/2,\pi/2)$
as expected for low doping. More relevant here, the single dispersion
curve evolves into a more complex structure: a) the $\omega>0$ part
not accessible by ARPES reveals a well-defined dispersion of weakly
damped QP, b) the renormalized band with small QP velocity remains
well defined close to the Fermi surface, i.e. at ${\bf k} \sim
(\pi/2,\pi/2)$, or even extends nearly to $(0,0)$ at low $T \to 0$,
c) less coherent band-like feature corresponding roughly to the bottom
of the unrenormalized band is well developed close to the zone center,
${\bf k} \sim (0,0)$. For $T<J$ both bands coexist at least at ${\bf k} \sim (\pi/4,\pi/4)$.

All observed features are present also
in $(\pi,0)-(0,0)$ direction as clearly seen in Fig.~1. However, the
difference appears at higher $T$ where the entire band in this direction
is positioned well below
the chemical potential and does not experience the waterfall effect
anymore. This is in agreement with the evolution of the band towards the
usual although still renormalized tight-binding dispersion.

\begin{figure}[htb]  
\centering 
\epsfig{file=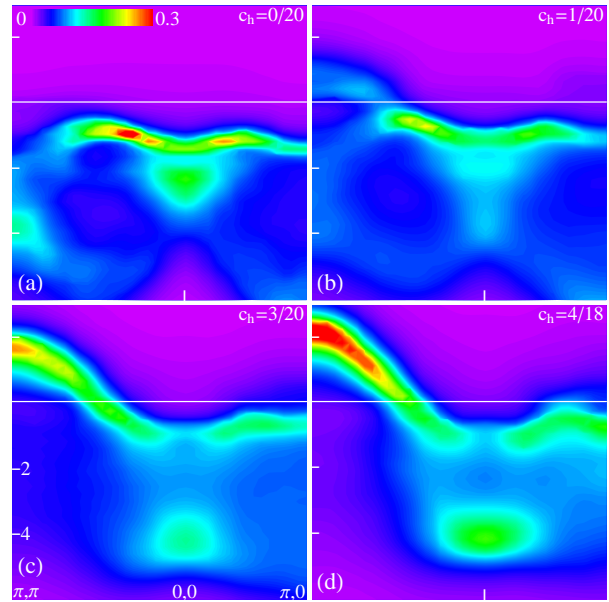,width=85mm,angle=0} 
\caption{(Color online) Weight map of $A({\bf k},\omega)$ vs. 
{\bf k} along symmetry lines for fixed $T/t=0.2$ and different hole
dopings $c_h$.}
\label{fig2}   
\end{figure}

Quite analogous behavior can be followed at fixed low $T$ as a
function of doping. We show results obtained using FTLM for systems
with $N=18,20$ within a broad doping range $c_h=0/20 - 4/18$.
Note that for the undoped system, $c_h=0$, the absolute position of the chemical potential is not well
defined within the $t$-$J$ model. In the latter case in Fig.~2a,
one can again recognize very well pronounced renormalized QP band
reaching the zone center, while the bottom band is
very incoherent. Both bands coexist in $(\pi/4,\pi/4)-(0,0)$ and
$(\pi/4,0)-(0,0)$ regions.
With increasing $c_h$ the renormalized QP band remains
well defined near the Fermi surface at ${\bf k} \sim (\pi/2,\pi/2)$, but
dissolves towards the zone center. At the same time the bottom band
starts to move away from the renormalized band and stays well pronounced near ${\bf k} \sim (0,0)$.
Both bands are connected with
the waterfall-like drop of low intensity. While the whole effective
bandwidth is weakly reduced $\Delta \omega \sim 6~t <8~t$ relative to
a tight-binding band, the bottom of the band at ${\bf k}=(0,0)$ is
deeper than expected from the tight-binding dispersion in the case of
$c_h=3/20, 4/18$. This is consistent with experimental observations \cite{meev}. 

The origin of the HEK can be best analyzed and understood by expressing
the single-particle Green's function corresponding to $A({\bf
k},\omega)=-\mathrm{Im}G({\bf k},\omega)$ in terms of the self energy
$\Sigma({\bf k},\omega)$,
\begin{equation}
G({\bf k},\omega)= \frac{\alpha}{\omega -\zeta_{\bf k} -
\Sigma({\bf k},\omega) }. \label{eq2}
\end{equation} 
The model, Eq.(\ref{eq1}), defined with projected fermionic operators
requires a nonstandard normalization $\alpha$ as well as a
nontrivial 'free' term $\zeta_{\bf k}$ representing the first
frequency moment of the $A({\bf k},\omega)$ \cite{prel}. Within the paramagnetic
metal with $\langle {\bf S}_i \rangle=0$ one can express explicitly
$\alpha=(1+c_h)/2$ and
\begin{equation}
\zeta_{\bf k} = \bar \zeta - 4 \sum_j r_j t_j \gamma_j({\bf k}), 
\quad r_j = \alpha + \frac{1}{\alpha} \langle {\bf S}_0 \cdot 
{\bf S}_j \rangle,  \label{eq3}
\end{equation} 
where $t_j, j=1,3$ represent hopping parameters $t,t',t''$,
respectively, which are renormalized with $r_j$ that involve local spin
correlations $\langle {\bf S}_0 \cdot {\bf S}_j \rangle$.
The tight-binding band dispersions corresponding to $t_i$
are then $\gamma_1({\bf k})= (\cos k_x +\cos k_y)/2$, $\gamma_2({\bf
k}) = \cos k_x \cos k_y$ and $\gamma_3({\bf k})= (\cos 2k_x +\cos
2k_y)/2$. The above expression, Eq.(\ref{eq2}), in terms of $\alpha$ and $\zeta_{\bf k}$ leads to
properly analytically behaved $\Sigma({\bf k},\omega \to \pm \infty)
\propto 1/\omega$.

Following Eq.(\ref{eq2}) we extract $\Sigma''({\bf k},\omega)$
provided that $A({\bf k},\omega)$ are smooth enough which is for
available systems typically the case 
for $T/t>0.1$.  In Fig.~3 we present results for
$\Sigma''({\bf k},\omega)$ corresponding to spectra in Figs.~1 at
$c_h=2/20$ and various $T/t$, but fixed ${\bf k}=(\pi/4,\pi/4)$ chosen
to represent the location of the HEK. It should be noted that
$\Sigma''({\bf k},\omega)$ is not crucially dependent on ${\bf k}$
(ignoring here more delicate phenomena as the pseudogap \cite{zeml1} ),
at least not inside the Fermi volume so results in Fig.~3 are
representative for all ${\bf k}$ relevant for effective ARPES bands.

Several characteristic properties of the QP damping recognized already
in previous studies \cite{jprev,zeml1,prel} can be deduced from
Fig.~3. a) The damping function $\Sigma''({\bf k},\omega)$ is very
asymmetric with respect to the Fermi energy $\omega=0$.  For the
hole-doped case discussed here the damping is large only for
$\omega<0$ corresponding to ARPES. b) As one expects in a metal we find
$\Sigma''({\bf k},\omega =0) \to 0$ (or at least decreasing) at low $T
\to 0$, a prerequisite for a well defined Fermi surface. c) Within
quite a large regime $-2t <\omega <0$ we recover at low $T$ well known
marginal variation $-\Sigma''({\bf k},\omega) \propto |\omega|$
\cite{varm,jprev}, while only at large $\omega<-3~t$ the damping
decreases and loses intensity. e) Increasing $T$ mainly influences the
behavior close to $\omega \sim 0$ filling the dip and increasing
$|\Sigma''({\bf k},\omega \sim 0)|$, at the same time making
$\Sigma''({\bf k},\omega)$ more featureless.

\begin{figure}[htb]  
\centering 
\epsfig{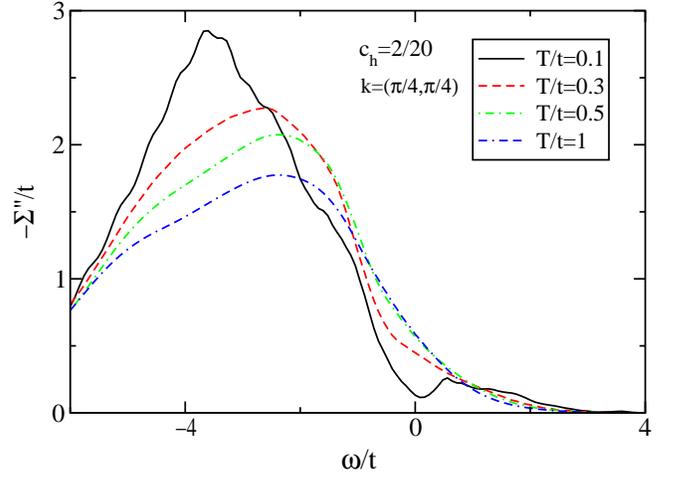} 
\caption{(Color online) Damping function $-\Sigma''({\bf k},\omega)$ corresponding to
Fig.~1 for $c_h=0.1$, ${\bf k}=(\pi/4,\pi/4)$ and various $T/t=0.1-1$.}
\label{fig3}   
\end{figure} 

Clearly, the strength and the form of $\Sigma''({\bf k},\omega)$
determines the anomalous dispersion $\omega_{\bf k}$ via
the pole location
\begin{equation}
\omega_{\bf k} - \zeta_{\bf k} + \frac{1}{\pi}\int d\omega' 
\frac{\Sigma''({\bf k},\omega')}{\omega_{\bf k}-\omega'}=0. 
\label{eq4}
\end{equation}

The relevant quantity to estimate the influence of $\Sigma''({\bf
k},\omega)$ on the dispersion $\omega_{\bf k}$ is the intensity
$\eta^2_{\bf k}= -\int \Sigma''({\bf k},\omega) d\omega/\pi$.  We
notice that at low doping $\eta_{\bf k}$ is not strongly dependent
either on ${\bf k}$, $c_h$ nor on $T$. In the range of interest
corresponding to Figs.~1-3 we find $\eta_{\bf k}^2\sim 3-4~t^2$. It
should be noted that the origin of large $\eta_{\bf k}$ is here
entirely in strong correlations, i.e., in the incoherent motion of a
particle (hole) in a spin background with singly occupied sites. Such
physics can be well captured by, e.g., a retraceable path approximation
\cite{brin} where one gets $\eta^2 = 4 t^2 $, very close to our
numerical results.

Since $\zeta_{\bf k}$ in Eq.(\ref{eq3}) produces only a regular although
renormalized tight-binding dispersion, the anomalous effective
dispersion emerges from $\Sigma'({\bf k},\omega)$. Due to large
$\eta_{\bf k}$ and a restricted range $-6~t< \omega <0$ of appreciable
$|\Sigma''({\bf k},\omega)|$, $\Sigma'({\bf k},\omega)$ leads to a
substantial change of the dispersion in this $\omega$ regime.
At low doping and $T<J$ it induces in combination with a narrow
$\zeta_{\bf k}$ a coexistence of renormalized QP band and the bottom band at ${\bf
  k}<(\pi/4,\pi/4)$. The latter one is
quite incoherent due to large $|\Sigma''({\bf k},\omega)|$ in
$\omega<0$ region. On the other hand at $\omega \sim 0$ one has
$\Sigma''({\bf k},\omega) \to 0$ which allows for a well defined
renormalized QP band near the Fermi surface.

The effect of $T>0$ is to broaden $\Sigma''({\bf k},\omega)$ and to
increase QP damping at $\omega \to 0$. Then, $\Sigma'({\bf k},\omega)$
shows less structure and the renormalized QP peak at low $\omega$
transforms with increasing $T$ into a single effective band.
However, due to $T$-independent $\eta_{{\bf k}}$ the structure of $\Sigma'({\bf
  k},\omega)$ remains strong enough to keep the waterfall drop
up to very high $T$.
Note that even for larger $T$ as in Figs.~1c,d an effective
dispersion following Eq.(\ref{eq3}) remains renormalized by $r_j \sim \alpha$ although
the band becomes wider as spin correlations loose intensity for $T>J$. Analogous are
phenomena at larger doping except that $\Sigma''({\bf k},\omega)$
generally decreases with $c_h$.

To illustrate that above features are essential and sufficient to
reproduce the HEK and the waterfall we compare numerical results in Fig.~1 with
a simplified model of $|\Sigma''({\bf k},\omega)|$ assuming: a)
$\Sigma''(\omega)$ is local, i.e., ${\bf k}$ independent, b) at $T=0$
it follows marginal behavior (linear in $\omega$) for
$-\epsilon_a<\omega<0$ \cite{varm}, c) for larger binding energies
$-\epsilon_b<\omega<-\epsilon_a$ it decreases linearly to zero, d) the
effect of $T>0$ is to convolute $\Sigma''(\omega,T=0)$ with usual
thermodynamic factor $f(\omega)[1-f(\omega)]$ where $f(\omega)$ is the
Fermi-Dirac distribution.

For results presented in Fig.~4 we fix
$\epsilon_a=t, \epsilon_b=6~t, \eta=2t$ and vary $\zeta_{\bf k}$
through $r_1=0.35, 0.5$ for $T=0, 0.75$, respectively,
while $r_{2,3}=\alpha$. We present in Fig.~4 the
$T$-dependence of $A({\bf k},\omega<0)$ with ${\bf k}$ along the zone
diagonal. It is well visible how the two-band structure at low $T$
transforms into a rather regular but broad
single band with persistent waterfall even at very high $T\sim t$. 

\begin{figure}[htb]  
\centering 
\epsfig{file=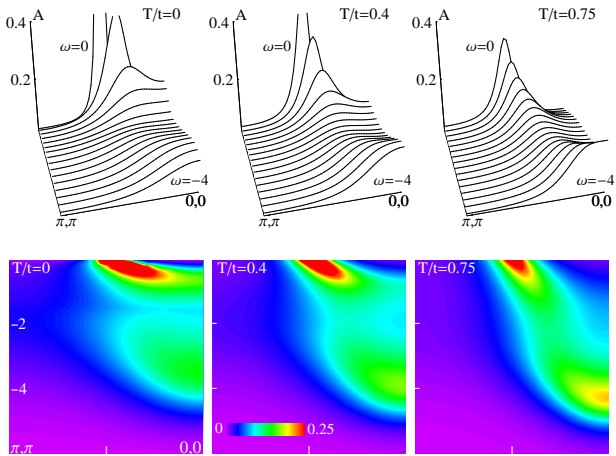,width=85mm,angle=0} 
\caption{$A({\bf k},\omega)$ along the zone diagonal calculated from a
simplified model and various $T/t=0, 0.4$ and $0.75$.}
\label{fig4}   
\end{figure} 

In conclusion, we have shown that the prototype model as the extended
$t$-$J$ model incorporates the physics of the HEK as well as the waterfall as
observed in numerous recent ARPES studies of hole-doped cuprates.
While at low $c_h$ and low $T$ the spectra typically reveal a
coexistence of a narrower renormalized QP band and an incoherent
bottom band most pronounced at ${\bf k}\sim (0,0)$ the structure evolves
with increasing either $c_h$ or $T$ into a single waterfall-like band
which persists up to very high $T\sim t$ or in the overdoped regime.
 
The origin of the anomalous dispersion is according to our analysis
entirely due to presence of strong correlations, as incorporated
already in the incoherent hole motion in a correlated insulator, as
given within the Brinkman-Rice scenario \cite{brin}. Such a conclusion offers also
the explanation why the waterfall phenomenon persists up to high $T>J$
and in a very broad range of hole doping $c_h$. Our results also
indicate that explanations in terms of specific low-$T$ features as
the AFM long range order \cite{mano} or AFM fluctuations \cite{macr,marw},
quantum critical point \cite{zhu} might be too narrow. In addition,
similar waterfall behavior can be observed also in 1D $t$-$J$ model at
high $T>J$ \cite{zeml3}. However, instead of an
incoherent bottom band a coherent holon branch
emerges with reducing $T<J$. This is different from the present case
in 2D.

There are also some predictions relevant for ARPES experiments
emerging from our analysis. In particular, ARPES spectra of
electron-doped cuprates should correspond to $\omega >0$ spectra
of hole-doped cuprates (although with opposite $t'$ and $t''$) as
already commented in \cite{zeml2}. From the large
asymmetry in $\omega$ as seen in presented results it follows that one cannot
expect the HEK and waterfall in ARPES results of electron-doped cuprates.
Further, our results predict an evolution of the anomalous
dispersion with increasing $T$ and $c_h$, nevertheless the waterfall
features should persist up to very high $T>J$ as well as in the overdoped
regime.

This work was supported by the Slovenian Research Agency under grant
PI-0044. T.T. acknowledges supports from the Next Generation
Supercomputing Project of Nanoscience Program, CREST, and Grant-in-Aid
for Scientific Research form MEXT, Japan.


\begin{thebibliography}{99}                                          
\bibitem{dama} A.\ Damascelli, Z.\ Hussain, and Z.-X.\ Shen, Rev.\
Mod.\ Phys. \textbf{75}, 473 (2003).
\bibitem{ronn} F.\ Ronning {\it et al.}, Phys.\ Rev.\ B \textbf{71}, 
094518 (2005).
\bibitem{meev} W.\ Meevasana {\it et al.}, Phys.\ Rev.\ B \textbf{75}, 
174506 (2007). 
\bibitem{graf} J.\ Graf {\it et al.}, Phys.\ Rev.\ Lett.\ \textbf{98}, 
067004 (2007).
\bibitem{vall} T.\ Valla {\it et al.}, cond-mat/0610249.
\bibitem{chan} J.\ Chang {\it et al.}, cond-mat/0610880.
\bibitem{pan} Z.\-H.\ Pan {\it et al.}, cond-mat/0610442.
\bibitem{inos} D.\ S.\ Inosov {\it et al.}, cond-mat/0703223.
\bibitem{xie} B.\ P.\ Xie {\it et al.}, Phys.\ Rev.\ Lett.\ \textbf{98}, 
147001 (2007).
\bibitem{park} S.\ R.\ Park {\it et al.}, Phys.\ Rev.\ B \textbf{75}, 
060501(R) (2007). 
\bibitem{kim} B.\ J.\ Kim {\it et al.}, Nature Phys. \textbf{2}, 387 (2006).
\bibitem{mano} E.\ Manousakis, Phys.\ Lett.\ A \textbf{362}, 
86 (2007). 
\bibitem{wang} Q.-H.\ Wang, F.\ Tan, and Y.\ Wan, cond-mat/0610491.
\bibitem{zhu} L.\ Zhu, V.\ Aji, A.\ Shekhter, and C.\ M.\ Varma,
cond-mat/0702187.  
\bibitem{bycz} K.\ Byczuk, M.\ Kollar, K.\ Held, Y.\-F.\ Yang, 
I.\ A.\ Nekrasov, Th.\ Pruschke, and D.\ Vollhardt, Nature Phys.
\textbf{3}, 168 (2007).
\bibitem{macr} A.\ Macridin, M.\ Jarrell, Th.\ Maier, and 
D.\ J.\ Scalapino, cond-mat/0701429.
\bibitem{marw} R.\ S.\ Markiewicz {\it et al.}, cond-mat/0701524.
\bibitem{tohy1} T.\ Tohyama and S.\ Maekawa, Phys.\ Rev.\ B
\textbf{64}, 212505 (2001).
\bibitem{tohy2} T.\ Tohyama, Phys.\ Rev.\ B \textbf{70}, 174517 (2004).
\bibitem{jprev} for a review see J.\ Jakli\v c and P.\ Prelov\v sek,
Adv.\ Phys.\ \textbf{49}, 1 (2000).
\bibitem{zeml1} M.\ M.\ Zemlji\v c and P.\ Prelov\v sek,
Phys.\ Rev.\ B \textbf{75}, 104514 (2007).
\bibitem{prel} P.\ Prelov\v sek and A.\ Ram\v sak,  Phys.\ Rev.\ B 
\textbf{63}, 180506(R) (2001); Phys.\ Rev.\ B 
\textbf{65}, 174529 (2002).
\bibitem{varm} C.M. Varma, P.B. Littlewood, S. Schmitt-Rink,
E. Abrahams, and A.E. Ruckenstein, Phys. Rev. Lett. {\bf 63}, 1996
(1989).
\bibitem{brin} R.\ Brinkman and T.\ M. Rice, Phys.\ Rev.\ B
\textbf{2}, 1324 (1970).
\bibitem{zeml3} M.\ M.\ Zemlji\v c, P.\ Prelov\v sek, and T.\ Tohyama,
unpublished.
\bibitem{zeml2} M.\ M.\ Zemlji\v c, P.\ Prelov\v sek, and T.\ Tohyama,
cond-mat/0702644.

\end{thebibliography}
\end{document}